# MAPSS, a Multi-Aspect Partner and Service Selection Method

Zbigniew Paszkiewicz and Willy Picard

Department of Information Technology, Poznań University of Economics
ul. Mansfelda 4, 60-854 Poznań, Poland
{zpasz, picard}@kti.ue.poznan.pl

**Abstract.** In Service-Oriented Virtual Organization Breeding Environments (SOVOBEs), services performed by people, organizations and information systems are composed in potentially complex business processes performed by a set of partners. In a SOVOBE, the success of a virtual organization depends largely on the partner and service selection process, which determines the composition of services performed by the VO partners. In this paper requirements for a partner and service selection method for SOVOBEs are defined and a novel *Multi-Aspect Partner and Service Selection* method, *MAPSS*, is presented. The MAPSS method allows a VO planner to select appropriate services and partners based on their competences and their relations with other services/partners. The MAPSS method relies on a genetic algorithm to select the most appropriate set of partners and services.

## 1 Introduction

A *Virtual Organization* (VO) is an operational structure consisting of different organizational entities and created for a specific business purpose, to address a specific business opportunit*y*. The success of a VO strongly depends on all participating organizations being capable of cooperating as a single unit, which usually implies an appropriate choice of partners.

The concept of *Virtual Organization Breeding Environment* (VOBE, sometimes referred in the literature as VBE) has been proposed to facilitate, among others, the VO creation process. A VOBE is "an association of organizations with the main goal of increasing preparedness of its members towards collaboration in potential virtual organizations" [1]. A VOBE provides a set of tools and data sources that may be used in partner and service selection process, including competence repositories, negotiation tools, history of collaboration etc.

As a valuable approach for the architecture and implementations of VOBEs and integration of cooperating organizations, the Service-Oriented Architecture (SOA) [2] has been suggested [3]. In this paper, only *Service-oriented Virtual Organization Breeding Environments* (SOVOBEs) are taken into account. In a SOVOBE, VOBE and VO operations are based on services performed by people, organizations and information systems, composed in potentially complex business processes.

Selection of services is usually separated from the concept of selection of partners. While service selection is a concept investigated by the distributed systems community (especially in terms of SOA), partner selection is a subject of interest of mainly the collaborative networked organization (CNOs) community. Nevertheless, in most methods presented in the works from both communities, it is possible to distinguish two elements: first, an *information model* captures and structures information about artifacts based on which the selection is performed. Second, various *selection techniques* have been proposed, focusing on the selection context, scope of the selection and selection strategy.

*Partner selection* is strongly connected with the idea of competence modelling [4]. The research works on competence modelling aim at providing a structural description of organizations with a special emphasis on competences, profiles, capacities, resources etc. [5, 6]. The concept of competence modelling is now developing rapidly and some works referring to partner selection based on organization profiles have been published [6, 7]. In a competence-based approach, inclusion of service characteristics is marginal [5] or not present at all [6].

In [8, 9], the *selection of services* is based on an information model consisting of service descriptions. A service description usually includes a wide range of technical, functional, non-functional and business characteristics of a given service [10]. In [11, 12] opinions concerning services and user feedback are taken into account. However, these works do not take advantage of the concepts elaborated in the area of competence modelling so proposed descriptions of service providers are often not structured, consisting of a simple list of attributes. Moreover, while the importance of social aspects in SOA has been noted recently [13, 14, 15], existing approaches are not mature and models for social relations and modelling requirements based on these relations are still to be developed. As a example, Ding and al. [16] have proposed a simulation-optimization approach using genetic search for supplier selection, integrating performance estimation, social aspects and genetic algorithm. However, the social relation model encompasses only a simple social model for supply chains limited to only one relation type, i.e. customer-supplier.

A number of selection strategies have been proposed for partner selection. The conclusion of the comparison of various popular approaches presented in [8] and [17] is that genetic algorithms are the most popular approach.

The main contribution of this paper is *MAPSS*, a novel method for *M*ulti-*A*spect *P*artner and *S*ervice *S*election in SOVOBEs supporting social aspects, competences of VO participants and performance characteristic. The proposed method has been implemented in the *ErGo* system within the ITSOA project [18].

This paper is organized as follows. In Section 2, basic notions related with collaboration in SOVOBEs are presented. In Section 3, the main requirements for the selection method are presented. In Section 4, a general outline of the method is described. The proposed information model is shortly described in Section 5 while the selection technique is the subject of Section 6. Finally, Section 7 concludes the paper.

## 2  Basic notions

*A business process* is "a set of one or more linked procedures or activities which collectively realise a business objective or policy goal, normally within the context of an organisational structure defining functional roles and relationships" [19].

A *business process definition* consists traditionally of "a network of activities and their relationships, criteria to indicate the start and termination of the process, and information about the individual activities, such as participants, associated IT applications and data, etc." [19]. The set of activities and their relations is referred to as *business process structure*. An aim of each partner and service selection method for VOs is to identify process participants and services that can perform activities identified in a process structure.

In the presented approach, each activity is performed by a *process participant* consuming a *service* provided by a *service provider*. A *process element* may refer to a process participant, a service or a service provider. Process participants and service providers are called *partners*. A *role* is a set of requirements that a potential partner or service has to satisfy to be assigned to a particular process element. Roles are in M:N relation with process elements.

Among requirements defining a role, social requirements concern relations with other roles. Examples of relations are past cooperation, recognition, former financial exchange, etc. The set of roles assigned to process elements and the relations among these roles are referred as a *social network schema*. A *social protocol* is a business process structure supplemented with social network schema.

## 3  Requirements for a partner and service selection method in SOVOBEs

The following requirements have been defined for a partner and service selection method:
- human control over the process – in context of complex business processes selection can hardly be fully automated;
- support for social aspects,
- support for competence descriptions,
- requirement-based approach – verification of social and non-social requirements coming from multiple sources (SOVOBE, VO planner, potential VO customers, etc.) defined for various elements of VO (process, partner, etc.)
- definition of VO planner's preferences,
- multi-variant analysis – evaluation of various partner and service compositions to maximize requirements satisfaction;
- indicator-based verification of VO network structure and performance characteristics, following a reference model proposed in [20].

## 4 Overview of the *MAPSS* method

The *MAPSS* method consists in five phases and follows the general selection method guidelines presented in [21]:
1. definition of VO specification – definition of requirements and associated preferences (importance, acceptable level of satisfaction etc.);
2. selection of partners and services for roles – selection of candidate elements; the output of this phase is a set of partners or services for each role defined in the social protocol;
3. VO variant generation – optimization of elements composition according to some fitness criteria defined by VO planner, this phase includes comparison of best possible VO variants; as an output a sorted set of variants is generated – a level of requirements satisfaction according to VO planner preferences is used as a sorting criteria;
4. performance evaluation – assignment of selected elements to process activities and validation of performance requirements;
5. VO inception – registration of the new VO in competence and service repository.

In every phase, human action may lead to requirements redefinition, preference modification, repetition of a steps, and reconfiguration of used supporting tools.

## 5 MAPSS information model

### 5.1 Competence and service description modules

In the *MAPSS* method, the description of services is extended with additional information concerning the organization providing these services (service providers). The additional description concerning the service providers takes the form of a competence description, based on the 4-C model [5], based on four key concepts: competence, capability, cost, and conspicuity. The competence and service description modules provide the following features used by *MAPSS*:
- structured description of organization competences and services,
- competence-based search of organizations based on submitted criteria consisting in required competences or other aspects described in model,
- evaluation of the conformance of an organization to a set of defined competence-based requirements,
- service description search based on submitted criteria consisting of required values of fields in service description,
- evaluation of the conformance of a service to a set of service description requirements.

### 5.2 Social network module

A social network is implemented as a graph of *nodes*, which may be connected by *relations*. The social network module provides information about the existence of relations among organizations and services registered in a SOVOBE. Relations in a social network are typed and contain a set of attributes. The social network module provides the functionality of verification of relations among organization. The module validates the level of conformance of a set of organizations to a set of social requirements.

### 5.3 Indicator module

In SOVOBEs, various services, provided either by the SOVOBE itself or SOVOBE members, may be considered as data sources which may be aggregated into indicators [20]. Indicators may allow a VO planner for definition of complex requirements involving various aspects, e.g. competences and social relations, of the planned VO in a single indicator. An important application of indicators is the verification of information from one data source by other data sources, e.g. information provided by organization in competence description may be confronted with the collaboration history stored in SOVOBE bag of assets.

The indicator module provides the functionality of indicator definition, storage and reuse.

### 5.4 Monitoring module

The *MAPSS* method is accompanied by a monitoring mechanism. The monitoring mechanism is based on the *Observer* design pattern [22]. A VO planner, being the Observer, may define indicators referring to objects (i.e. competence) and values (i.e. resource amount) existing in SOVOBE or events (i.e. registration of new organization with particular competence, update of service parameters, changes in the social network). When events or changes concerning objects or values occur, indicators are recalculated. If the indicator reaches a predefined alarm level, the VO planner is informed and may potentially change the set of services that has already been selected.

## 6  *MAPSS* selection technique

The *MAPSS* selection technique consists of 5 phases as presented in Section 4.

### 6.1 Phase 1: Definition of VO specification

A *virtual organization specification* consists of:
- a set of *requirements* concerning various *aspects* of the virtual organization;

- *preferences* – optimal, reject values and weights referring to requirements defined for selection process;
- *fitness functions* – global fitness functions, and multi-attribute utility functions used in the second and third phase;
- acceptable requirement conformance levels to limit the size of various sets of process elements generated during the execution of *MAPSS*.

*Aspects* of VO specification include: process elements, process, subsets of partners and subsets of services. It is possible to distinguish three types of requirements:
- *roles* referring to elements, e.g. localization, required set of competences, level of available resources, response time of a service, cost of the service; requirements are based on the competence and service description model;
- *social requirements* referring to structure of relations among organizations and services, e.g. past cooperation, recognition, use of service, recommendation; the main source of social requirements is a social network schema, but indicators may also provide the VO planner with means to define social requirements.
- *performance requirements* referring to evaluation of composition of partners and services in a process in terms of performance, e.g. maximum time of process duration or maximum time of sub-process response time.

Table 1 presents the structure of VO specification and the potential use of different requirements types in various phases of MAPSS.

**Table 1.** VO specification structure.

| Aspects | Roles | Social req. | Performance req. |
|---|---|---|---|
| Partner | Phase 2 | | - |
| Service | Phase 2 | | - |
| Subset of partners | - | Phase 3 | - |
| Subset of services | - | Phase 3 | Phase 4 |
| Process | - | - | Phase 4 |

### 6.2 Phase 2: Selection of partners and services for roles

Based on role requirements, a set of services or organizations are selected for each role. Identified elements may be sorted and filtered out according to the level of conformance to the associated role requirements.

### 6.3 Phase 3: Generation of VO variants

The number of potential VOs that may be constituted with services and organizations identified in phase 2 is usually high. The goal of the third phase is to search in the usually large domain of potential VOs a sorted list of VO *variants* ranked according to a *fitness function*. A VO variant is a set of services and organizations assigned to the roles defined in phase 1 of the *MAPSS* method

In phase 3, a genetic algorithm is used for the determination of the best fitted VO variants, as illustrated in Figure13. The genome $g$ is an array of N items, *N* being the number of roles. Genes *re* are sets of partners or services selected in phase 2 and assigned to roles *r* from the genome *g*.

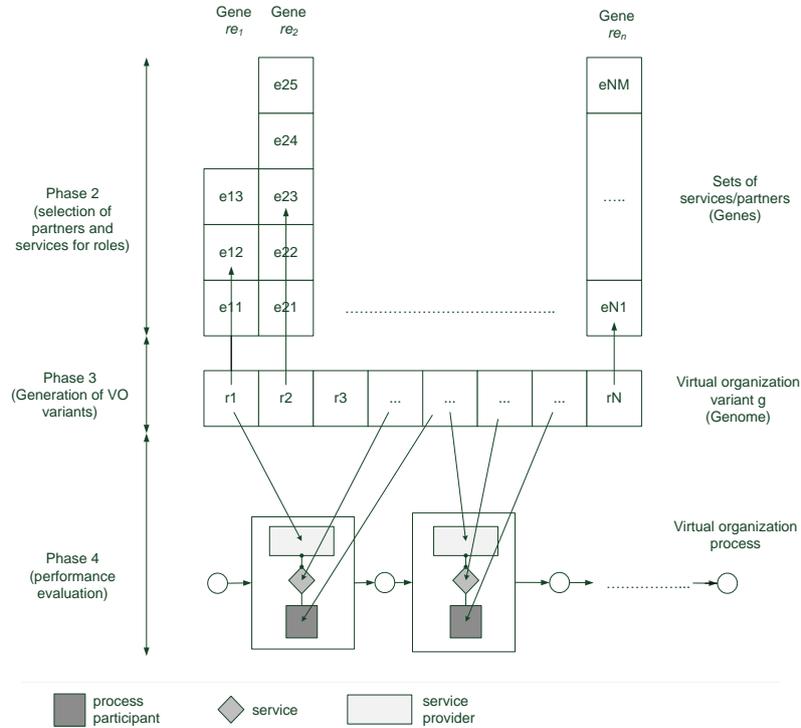

**Fig. 1. Phases 2-4 of MAPSS**

The crossover operator is the standard two-point crossover, while the mutation operator randomly selects a role (a position in the genome) and randomly replaces the corresponding partner or service with another one among those available from the set of elements assigned to this role. In this way the genetic algorithm creates many VO variants. Each variant is evaluated with the use of the fitness function defined by VO planner in VO specification. The *fitness function* used in this phase estimates the level of satisfaction of social requirements. The result of this phase is a sorted set of VO variants. The value of the fitness function calculated for each VO variant is used as sorting criteria. A *threshold value* defined with the VO planner in phase 1 is used to filter out the VO variants: VO variants for which the value of the fitness function is below the threshold value are not passed to phase 4.

### 6.4 Phase 4: Performance evaluation

The goal of phase 4 is to establish a *ranking* of VO variants selected in phase 3 according to a *fitness performance function*. The fitness performance function is defined by the VO planner, based on the performance requirements defined in phase 1. A fitness performance function may take into account various performance aspects, including operational performance, effectiveness, responsiveness, cost, social relations among VO elements.

The values of a fitness performance function may be vectors, where each vector component is associated with the evaluation of a particular performance aspect. A *ranking function* is responsible for sorting the values of the fitness performance function for the VO variants. As a consequence, various performance evaluations may be performed on a given set of VO variants, either with different fitness performance functions, or with difference ranking functions.

### 6.5 Phase 5: VO continuance

The VO variant proposed as the best ranked in phase 4 is registered as a VO in the SOVOBE, i.e. the VO is registered in the competence description module as a new organization in SOVOBE, having non-empty set of competences and non-empty set of services.

### 6.6 Technical implementation

The *MAPSS* method has been implemented as part of the *ErGo* system developed within the ITSOA project [18], in the Java programming language. The functionality provided by the *MAPSS* modules is exposed as OSGi services [23] within the Equinox container [24] and can be externalized as SOAP web services. Modularity of the system supports exchange of system modules, e.g. other implementation of competence and service description, and social network modules can be used.

## 7 Conclusions

The main contribution presented in this paper is a novel method for multi-aspect service selection, integrating competencies, social aspects and performance. While each of these topics have already been studied, the novelty of the proposed method lays in their combination.

The method proposes a structured approach to requirement definition in a form of VO specification. This approach distinguishes different requirement types, method phases for their use and selection technique for verification of their satisfaction. In particular social requirements are considered.

The potential involvement of the VO planner at various steps of the proposed method, combined with the possibility to loop back to formerly performed steps, meets the complex, often iterative, nature of partner and service selection. However, it results in a potentially undetermined duration of the method execution.

The important issue of privacy related with information about SOVOBE members is not taken into account in the proposed method. As a consequence, it is assumed that all information related with competencies, social relations and performance are publicly available, which is usually not wanted in real cases.

The method was implemented as a part of the *ErGo* system.

Among future works, the *MAPSS* method should be extended to encompass the collaborative nature of the partner and service selection process. In its current form, the *MAPSS* method is based on the assumption that a VO planner is the only responsible person/organization for the selection process.

Within the IT-SOA project [18], further development of the proposed method and its verification is planned with a pilot application in the construction sector.

**Acknowledgments.** This work has been partially supported by the Polish Ministry of Science and Higher Education within the European Regional Development Fund, Grant No. POIG.01.03.01-00-008/08.